\documentclass[pre,floatfix,showpacs,twocolumn]{revtex4}%
\usepackage{amsfonts}
\usepackage{amsmath}
\usepackage{amssymb}
\usepackage{graphicx}
\usepackage{color}
\usepackage{psfrag}
\usepackage{verbatim}
\usepackage[tight,sf]{subfigure}%
\providecommand{\U}[1]{\protect\rule{.1in}{.1in}}
\begin{document}
\title{Confotronic Dynamics of Tubular Lattices}
\author{Osman Kahraman$^{1}$, Herv\'e Mohrbach$^{1,2}$, Martin Michael M\"uller$^{1,2}$, Igor M. Kuli\'c$^{2}$}
\affiliation{$^{1}$Equipe BioPhysStat, ICPMB-FR CNRS 2843, Universit\'{e} de Lorraine; 1
boulevard Arago, 57070 Metz, France}
\affiliation{$^{2}$CNRS, Institut Charles Sadron; 23 rue du Loess BP 84047, 67034
Strasbourg, France}
\date{\today }

\begin{abstract}
Tubular lattices are ubiquitous in nature and technology. Microtubules and nanotubes of all kinds act as important pillars of biological
cells and the man-made nano-world. We show that
when prestress is introduced in such structures, localized conformational
quasiparticles emerge and govern the collective shape dynamics of the lattice.
When coupled \textit{via} cooperative interactions these quasiparticles form
larger-scale quasipolymer superstructures exhibiting collective dynamic modes
and giving rise to a hallmark behavior radically different from semiflexible beams.

\end{abstract}

\pacs{87.16.aj,82.35.Pq,87.15.-v}
\maketitle



\section{Introduction}

Tubular and cylindrical lattices are abundant in living nature and
give rise to important filamentous structures, such as bacterial
flagella and microtubules. Inspired by biology, nanotubes made
from various building blocks like carbon \cite{Iijima}, DNA
\cite{mahtieu} or amphiphilic molecules \cite{micelles,micelles2}
have been synthesized. Remarkably, in the presence of internal
prestress, almost all of these tubular objects can adopt
superhelical structures, \textit{i.e.}, tubes whose centerline
describes a large-scale helix in space. Superhelical carbon
nanotubes with micron-meter size pitch have been observed
\cite{Xang carbon} and their elastic properties probed
\cite{Volodin}. It is believed that their coiling results from the
periodic arrangement of defects (pairs of heptagon-pentagon rings)
in the hexagonal carbon lattice forming the wall of the tube
\cite{Dunlap carbon}. DNA nanotubes \cite{Douglas DNA} adsorbed on
a substrate resemble squeezed helices indicating that they assume
a three-dimensional superhelical structure under free conditions
\cite{Squeezed helix}. Supramolecular chiral helical nanofibers
made of self-assembling lipids have been synthetized. It was shown
that the torsional stress, created by the steric interactions of
the chemical groups at the surface of the tube, causes the
nanofibers to coil into a superhelix that minimizes the internal
prestress \cite{Li micelles}. In bacterial flagella and
microtubules the coexistence of several conformational states of
their individual constituents is in elastic conflict with their
lattice geometry, resulting in prestresses that can be minimized
by forming superhelical shapes \cite{Calladine,mt2010,mt2012}. The
protein monomer units of bacterial flagella are arranged in eleven
protofilaments parallel to the centerline that can switch from a
short to a longer state and can slide down relative to their
lateral neighbors creating local twist and curvature
\cite{Calladine}. In a similar manner, the wall of microtubules is
made of protofilaments built by the polymerization of tubulin
dimers. There are experimental indications that the tubulin dimers
can indeed switch between straight and curved conformational
states in the presence of taxol \cite{Amos} and display an
allosteric cooperative interaction along their protofilaments'
axes \cite{ElieCaille2007}. The integrity of the tubular lattice
can be maintained either by forcing all protofilaments in their
straight state or by creating a mixed phase with a cluster of
curved protofilaments while the rest of the protofilaments stays
in their straight state. When this phase is energetically more
favorable, the microtubule bends in the direction of the block
formed by the curved protofilaments. Since microtubules have
internal twist \cite{Chretien} (\textit{i.e.}, the protofilaments
are not parallel to the centerline of the tube but instead wind
around it), the resulting shape will be a superhelix whose pitch
is believed to be given by the internal twist
\cite{mt2010,mt2012}. Remarkably there is some evidence that, in
contrast to the bacterial flagella filament and all other
structures discussed,
microtubules are super-helices that are spontaneously and
permanently reshaping on experimental timescales: they change
their reference ground state due to thermal fluctuations. This
unusual collective movement was previously proposed and termed the
``wobbling mode'' \cite{mt2010,mt2012}.  

Notable phenomenological models for multistable helices have been 
developed in the past in order to describe transformations of bacterial flagella 
\cite{Powers,WadaNetz1,Friedrich}, coiled plant tendrils \cite{TendrillPerversion}, 
or whole microorganisms \cite{WadaNetz2,BistableHelices}.
The novel and rather unique feature of our present model, that we
will explore and illustrate here in depth, will be the
extraordinary dynamic behavior associated with the cooperative
``wobbling motion'' of the filament.
Although biofilaments are usually studied in the framework of beam
elasticity, microtubules can also be modelled as cylindrically
wrapped membranes since they are hollow. J\'{a}nosi \textit{et
al.}, for example, modelled microtubule walls as elastic sheets in
order to analyze their elastic properties \cite{Janosi1998}.
Inspired by the polymorphic tube model previously proposed by some of
the authors \cite{mt2010,mt2012,mtsolvable} we study a tubular system
that incorporates the idea of lattice confinement of bistable
units and cooperativity into an elastic sheet. Numerical
simulations and analytical models are combined here to provide an
in-depth intuitive understanding of this system. The prestress,
that will be built into our model, will give rise to a remarkable
phenomenon: Localized conformational deformations, that will behave
as quasiparticles, will emerge and govern the collective shape
dynamics of the lattice via elastically-mediated interactions.
When we switch on additional mechanical coupling terms in the
lattice, these quasiparticles will exhibit cooperative
interactions. The cooperativity will lead to the formation of
larger-scale ``quasipolymer'' superstructures that we will show to
exhibit rather unusual collective dynamic modes. The notion of
quasiparticles/-polymers is the most natural language to
quantitatively describe many new phenomena for which we will
collectively use the term \textquotedblleft confotronic
dynamics\textquotedblright.

In real biofilament systems this dynamics is usually inaccessible
to direct observation. However, as the internal confotronic modes
also govern the behavior of the centerline of the tube as a whole,
their existence can be experimentally inferred from the
observation of anomalous behaviors of the tube's centerline in
well chosen experiments. Among them, observing the dynamics of a
tube clamped at one end, turns out to be the experiment of choice.
We will see that a lot about the inner confotronic dynamics of
quasiparticles can be revealed from the external behavior of the
tube.

The paper starts with a description of the polymorphic tube model
in Sec.~\ref{sec:model}. In Sec.~\ref{sec:conformations}, the
notion of quasiparticles (called ``confoplexes'') is presented in
detail. We will see that these particles can form ordered
conformational superstructures on the lattice
 (called ``confostacks''). This order is shown to be responsible
for the formation of the broken symmetry superhelical tube state,
very much akin to the superhelices observed in microtubules and
bacterial flagella. Finally, in Sec.~\ref{sec:dynamics} the
confotronic dynamics of clamped polymorphic tubes is analyzed with
a particular focus on the remarkable collective mode (``wobbling
mode'') that emerges spontaneously in the system at finite
temperature.


\section{The Polymorphic Tube Model\label{sec:model}}

\begin{figure}[t]
\centering
\subfigure[][]{\label{fig:tube}\includegraphics[width=0.23\textwidth]{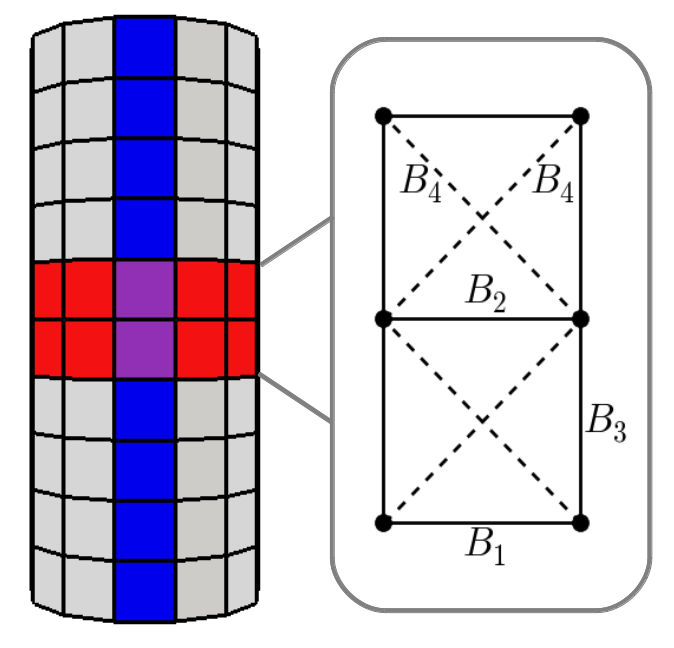}}
\hspace{0.75cm}
\subfigure[][]{\label{fig:dwell}\includegraphics[width=0.19\textwidth]{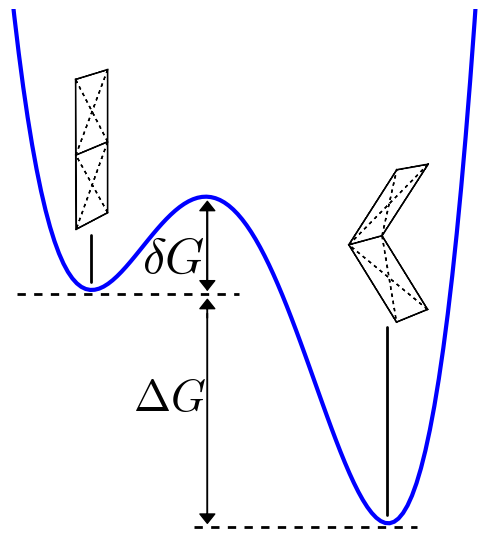}}
\newline%
\caption{(a) Construction of the polymorphic tube out of
rectangular subunits (dimers). One ring of lateral neighbors is
colored in red. One (proto-)filament formed by longitudinal
neighbors is colored in blue. A rectangular subunit is divided
into two squares by bonds $B_{2}$. Ghost diagonals $B_{4}$ are
added for shear rigidity. (b) Asymmetric double well potential for
the bonds $B_{2}$. $\Delta G$ denotes the energy difference
between the straight and the curved state. $\delta G$ is the
height of the barrier with respect to the straight state.
}%
\label{fig:MTsystem}%
\end{figure}

We construct a hollow polymorphic tube by discretizing its surface
as a mesh of rectangular subunits (see Fig.~\ref{fig:tube}).
Similarly to previous discrete models of (linear) elastic
membranes \cite{SeungNelson88,Janosi1998}, the elastic properties
of the tube will be enforced via stretching and bending energetic
penalties of the mesh. The polymorphic character of the tube is
implemented at the level of each subunit, decomposed into two
squares, which is multistable and can be either straight or curved
(see Fig.~\ref{fig:dwell}). Note that this choice of mesh is well
adapted to describe biofilaments such as microtubules. In this
picture, each rectangular, double-square subunit of the mesh
corresponds to a single tubulin dimer. The association of the
former along the vertical direction defines protofilaments (blue
in Fig.~\ref{fig:tube}) which form the tube in a way similar to
protofilaments forming the wall of a microtubule. In this work we
always consider $13$ such protofilaments alluding again to
microtubules. To stay as generic as possible, however, we will
omit many system-specific details like the internal twist and
other particularities of microtubules like the \textquotedblleft
seam\textquotedblright\ \cite{Janosi1998}. Such generic
simplifications, including symmetry, are necessary in order to
extract the physical ``gist'' of such systems, as we will see from
the plethora of phenomena emerging already in this simplified
lattice geometry.

Each dimer consists of internal and external bonds. The horizontal
internal bond $B_{2}$ separates the subunit into two squares and
forms a \emph{hinge} in the curved state of the dimer. The
internal bonds $B_{4}$, depicted by dashed lines in
Fig.~\ref{fig:tube}, control the shearability of each square and
ensure their planarity. The external bonds are shared with
neighboring dimers: two bonds of type $B_{1}$ along the vertical
and four of type $B_{3}$ along the horizontal direction.
 The elastic energy of the tube is a sum of contributions for
stretching and bending of the mesh, $E = E^\text{S}+E^\text{B}$.
Denoting $\mathbf{d}_{i}$ the vector associated to a bond of type
$B_{i}$, the total stretching energy of \textit{all} bonds of type
$B_i$ is given by
\begin{equation}
E_{i}^{\text{S}}=\frac{1}{2}\mu_{si}\sum_{\left\{  \mathbf{d}_{i}\right\}
}\left(  \left\vert \mathbf{d}_{i}\right\vert -d_{i}^{(0)}\right)
^{2},\,\,i=1,2,3,4
\label{eq:stretchingenergy}
\end{equation}
in the harmonic approximation. The sum runs over all the edge
vectors $\mathbf{d}_{i}$ of the tube. In
Eq.~(\ref{eq:stretchingenergy}), $d_{i}^{(0)}$ is the preferred
length of the bond $B_{i}$ when the tube is in its straight
cylindrical state (Fig.~\ref{fig:tube}) and $\mu_{si}$ is the
stretching rigidity. The stretching energy of the
whole mesh is thus: $E^\text{S}=\sum_{i=1}^4 E_{i}^{\text{S}}$.

Similarly, we associate a quadratic bending energy with all bonds of type
$B_{1}$, $B_{3}$, $B_{4}$:
\begin{equation}
E_{i}^{\text{BH}}=\frac{1}{2}\mu_{bi}\sum_{\left\{  s_{i}\right\}
}\left(  s_i-s_{i}^{(0)}\right)  ^{2},\,\,i=1,3,4
\end{equation}
where ${s}_{i}{=(\mathbf{n}^{a}\times\mathbf{n}^{b})\cdot\frac{\mathbf{d}_{i}}{|\mathbf{d}_{i}|}}$ is the sine of the angle between the outward normals
$\mathbf{n}$ of two adjacent triangles $a$ and $b$ at bond $\mathbf{d}_{i}$,
and $\mu_{bi}$ is the bending rigidity of this bond. The
constants $s_{i}^{(0)}$ are chosen such to enforce the straight
cylindrical state.
To allow the subunits to accommodate two stable conformations, we assign an
anharmonic bending potential $E_{2}^{\text{BAH}}$ for all bonds of type $B_{2}$:
\begin{equation}
E_{2}^{\text{BAH}}=\sum_{\left\{  s_{2}\right\}  }\left(  As_{2}^{4}+Bs_{2}^{3}+Cs_{2}^{2}\right)  \;,\label{eq:anharmonic}\end{equation}
where the coefficients $A$, $B$, and $C$ are all functions of the energy
difference $\Delta G$ and the barrier $\delta G$ and are chosen to favor the
curved state
(see Fig.~\ref{fig:dwell}). The total bending energy of the mesh is thus
$E^\text{B}=E_{1}^\text{BH}+E_{2}^\text{BAH}+E_{3}^\text{BH}+E_{4}^\text{BH}$.

In the following, the values of the stretching rigidities will be
expressed in units of $k_{B} T_0 / d^{2}$ and the bending
rigidities in units of $k_{B} T_0$, where $T_0$ is the room
temperature and $d=4\,$nm the size of a monomer (see
appendix~\ref{app:simulations} for all the parameter values). The
temperature $T$ in the simulations will be measured in units of
$T_0$.


\section{Confoplexes, confostacks and spontaneous symmetry breaking\label{sec:conformations}}

The rich conformational properties of the previously defined tube
model will now be analyzed with the help of numerical simulations
and phenomenological models. The large number of parameters of the
model is a serious limitation for exploring all possible
conformations in an attempt to build a complete phase diagram.
Instead, we reduce the number of independent parameters as much as
possible and look for interesting generic configurations and
behavior. In this spirit, we set
$\mu_{s}=\mu_{s1}=\mu_{s2}=\mu_{s3}$ and choose their value in the
typical range of the elastic constants of microtubules
\cite{Janosi1998} (see below). To facilitate the visual and
numerical detection of the characteristic behavior for the short
lattices practically accessible to our simulations, we have
deliberately chosen an intrinsic curvature of the protofilaments
much larger than for real microtubules. The other parameters are
adjusted to ensure the numerical stability of the mesh (see
appendix~\ref{app:simulations}).

We first discuss the results of the numerical simulations in which the
system is integrated in time with the Langevin Dynamics method (see again
appendix~\ref{app:simulations}). To understand the subtle competition between
the anharmonic potential~(\ref{eq:anharmonic}) and the elasticity of the
lattice, our study is built up in a hierachical manner: first the different
conformations of a single section of a tube are explored, from which, in a
second step, we can understand the behavior of longer tubes. Finally, long
tubes with additional cooperative interactions along the protofilaments are studied.


\subsection{Numerical simulations}

\subsubsection{A single section of the tube, the emergence of a confoplex}
\begin{figure}[t]
\centering
\includegraphics[width=0.35\textwidth]{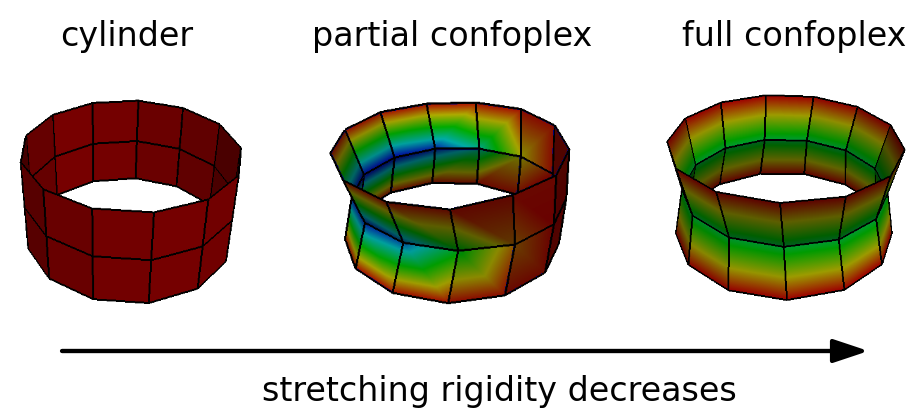}
\caption{Three configurations for one ring of dimers depending on the stretching
rigidity $\mu_{s}$: the straight cylinder, a partial confoplex, and a full
confoplex for which all dimers are in the curved state.
}%
\label{fig:dimerring}%
\end{figure}

We first consider the ground state (zero temperature limit) of a
single section of the tube. By varying the elastic stretching
constant $\mu_{s}$ we observe three different types of
configurations (see Fig.~\ref{fig:dimerring}): For sufficiently
large values of $\mu_{s}$ (at the order of $10^{5}$) the dimers
can not switch to their curved configuration and the ring
maintains its cylindrical form. However, when $\mu_{s}$ becomes
smaller ($2\times10^{4}$), all subunits adopt a curved
conformation, forming what we call a ``conformational complex''
or---more briefly---a \textit{confoplex}. The ring is still
cylindrically symmetric but its shape is catenoid-like. Very
interestingly, for intermediate values of $\mu_{s}$ (around
$4\times10^{4}$) a state with \textit{broken cylindrical symmetry}
is the minimum-energy configuration. In this \textit{partial
confoplex} conformation a cluster of neighboring dimers on one
side of the ring is switched to the curved state, thereby creating
a negatively curved scar which gradually decays towards the
opposite side of the tube. This causes the opposite wall to 
bulge slightly outwards.
\begin{figure}[t]
\centering
\subfigure[][]{\label{fig:2body}{\includegraphics[height=0.2\textheight]{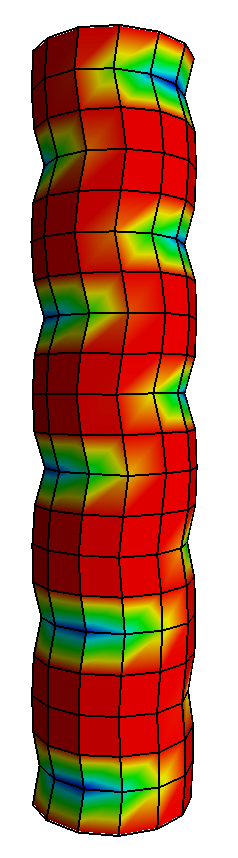}}}
\hspace{0.5cm}
\subfigure[][]{\label{fig:coop20}{\includegraphics[height=0.22\textheight]{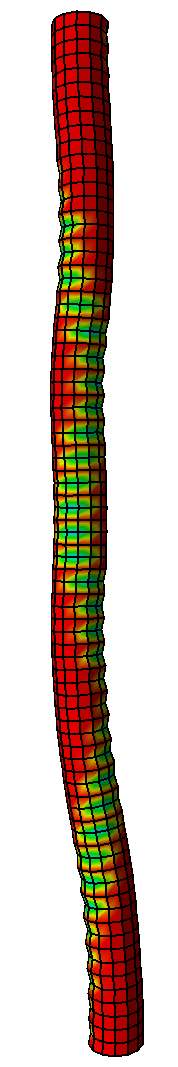}}}
\hspace{0.5cm}
\subfigure[][]{\label{fig:coop14}{\includegraphics[height=0.22\textheight]{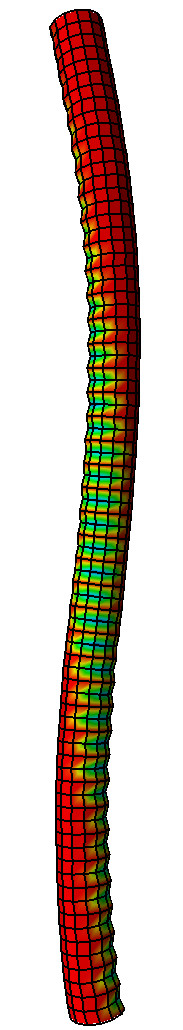}}}
\hspace{0.6cm}
\subfigure[][]{\label{fig:coop12}{\includegraphics[height=0.22\textheight]{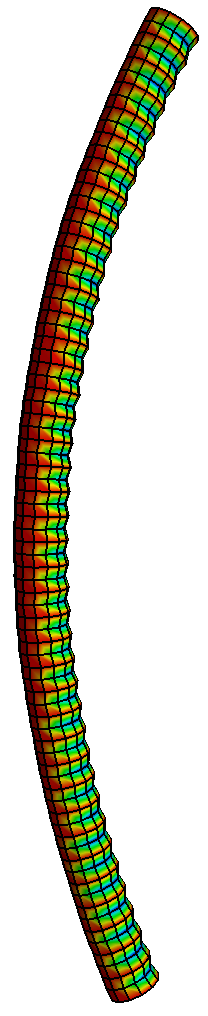}}}\caption{
(a) Repulsive partial confoplexes form a zigzag pattern on the tube ($\mu
_{b1}=300$, $T=3$). (b)--(d) Simulation snapshots for a system with fixed
cooperativity $\mu_{c}=50$ and $N=40$ segments at a temperature $T=3$. (b) The
repulsive interactions are dominant for large $\mu_{b1}=300$. (c) For weaker
repulsion ($\mu_{b1}=200$) a large-scale order emerges. (d) For $\mu_{b1}=0$,
the tube deflects into a C-like shape. }%
\label{fig:coop}%
\end{figure}


\subsubsection{Interacting confoplexes}

The emergence of full and partial confoplexes in the ring leads to
the interesting question of how single confoplexes interact with
each other in a larger lattice. In contrast to the single section, 
which does not possess any longitudinal neighbors, the
bending around the bonds of type $B_{1}$ will now have an
important effect. For the case of full confoplexes successive
sections meet at positive curvatures at the $B_{1}$ bonds. Hence,
for non-zero $\mu_{b1}$, stacking one full confoplex on top of a
second one will cost some energy; when the value of $\mu_{b1}$ is
high enough to dominate the other terms, the tube goes back to the
straight configuration.

For values of $\mu_{s}$ for which the tube is composed of partial
confoplexes, the deflection of each section is expected to be in
almost independent directions if $\mu_{b1}$ is zero. (In fact,
they are not completely uncorrelated due to the stretching energy
of the bonds $B_{1}$.) This almost random mutual orientation has
been confirmed by simulations at finite temperature (not shown).
However, for non-zero values of $\mu_{b1}$, (\textit{e.g.} for
$\mu _{b1}=300$), each section conserves its partial confoplex
configuration but neighboring confoplexes tend to align their
orientation opposite to each other, forming an \textquotedblleft
antiferromagnetic\textquotedblright-like zigzag pattern along the
tube (see Fig.~\ref{fig:2body} for an example at temperature T=3).
If we keep increasing $\mu_{b1}$, the confoplexes become gradually
smaller in size and finally vanish, similar to the
catenoid-to-cylinder transition found for the ring when $\mu_{s}$
is increasing.


\subsubsection{Cooperativity: confostacks, helices and spontaneous symmetry breaking}

The repulsive interaction of confoplexes implies that global
conformational superstructures of the tube are formed only if
neighboring confoplexes interact cooperatively via additional
interactions. Cooperative interactions among monomers in a
filament have been observed, for example, in single protofilaments
of microtubules \textit{in vitro} \cite{ElieCaille2007}. Motivated
by this observation, we implement the coupling between neighboring
dimers on a protofilament by adding a cooperative interaction
between the angles associated to the anharmonic bonds $B_{2}$:
\begin{equation}
E^{\text{CP}}=\frac{\mu_{c}}{2}\sum_{\left\{  s_{2}\right\}  }\left(  \left(
s_{2}-s_{2,\mathrm{next}}\right)  ^{2}+\left(  s_{2}-s_{2,\mathrm{prev}%
}\right)  ^{2}\right)  \;,
\end{equation}
where $\mu_{c}$ is the strength of the cooperative interaction
between a bond $B_{2}$ and its nearest neighbors (\textit{i.e.},
$B_{2}$-next and $B_{2}$-previous). In the presence of partial
confoplexes this term causes an attractive interaction between
them, opposing the repulsive one (due to the bending $\mu_{b1})$
and generating an interesting ``frustrated'' situation. The
attraction between neighboring partial confoplexes leads to a
stack of confoplexes forming a kind of ``quasipolymer''
superstructure arrangement on the lattice. This particular
internal state of the lattice, being an ordered stack of
confoplexes, will from now on be called a
\textit{``confostack''}. In a similar manner as the emergence of
the confoplex previously gave rise to the breaking of the cylindrical
symmetry on the \emph{single section scale}, the formation of the
ordered confostack is responsible for the spontaneous breaking of
the \emph{large scale} chiral symmetry of the tube
centerline in the three-dimensional space.

At finite temperature, thermal fluctuations tend to disorganize
the ordered confostack. In this case we observe the formation of
uncorrelated fluctuating domains of ordered confoplexes
distributed along the confostack. The characteristic size of such
a domain will be called the coherence length of the confostack and
denoted with $l_{c.}$ This coherence length shares some analogy to
the concept of persistence length of a semiflexible polymer. In
particular we can write $l_{c}=C/(k_{B}T)$ where $C$ (an unknown
function of the material parameters) can be seen as an effective
stiffness of the confostack. Note that an exact computation of
$l_{c}$ for a simplified model of microtubules can be found in
Ref.~\cite{mtsolvable}. We expect two extreme situations: For a tube of 
length $L$ such that $L\ll l_{c},$ the confostack is fully ordered and
the entire tube breaks the cylindrical symmetry. For $L\gg l_{c},$
the confostack is made up of a juxtaposition of uncorrelated
fluctuating domains of ordered confoplexes leading to a
statistically straight tube on average.

What is the fundamental shape of a single domain of an
ordered stack of confoplexes? The answer comes from the numerical
simulations. Figs.~\ref{fig:coop20}-\ref{fig:coop12} show a number
of snapshots of the typical configurations of the tube at finite
temperature for increasing values of the ratio
$\nu:=\mu_{c}/\mu_{b1}.$ In Fig.~\ref{fig:coop20} where $\nu$ is
small, we see that the confostack is made of short coherent
helices of random handedness. This corresponds to the situation $L\gg l_{c}%
,$ where the tube looks straight on long scales, since right- and
left-handed helices cancel each other out. The fundamental shape
of a domain of ordered confoplexes is thus a helix with a pitch
which depends on $\nu .$  The origin of this helicity will be
elucidated below in section~\ref{subsubsec:confostackformation}.

Increasing $\nu$ leads to a stronger alignment of the
partial confoplexes, i.e., an increase of the pitch of the
confostack. When the coherence length is larger than $L$ but the
pitch is still smaller than $L$, the switched dimers form a
coherent helical confostack, deflecting the centerline of the tube
into a superhelix in space as shown in Fig.~\ref{fig:coop14}. Note
that the superhelices of either handedness appear spontaneously
within a completely symmetric lattice. This can be seen as an
indication that real microtubules might form superhelices with
finite pitches by a similar spontaneous symmetry breaking
mechanism, even in the absence of an explicit internal lattice
twist \cite{mt2010,mt2012}. For very large $\nu $ the
cooperativity is strong enough for the pitch of the helical
confostack to become larger than $L$. Fig.~\ref{fig:coop12} shows
such a situation with $L\ll l_{c}$. In this case the tube forms a
circular arc with an untwisted confostack living on it.


\subsection{Phenomenological modelling}
After outlining the empirical observations of various
interesting phenomena from the Langevin simulations in the
previous section, in this section we seek to analytically and
phenomenologically explain the observations. In the first
subsection we will explore how the very existence of full/partial
confoplexes can be qualitatively understood within a simple
analytical model. Following that, in the second subsection, we
outline a simple phenomenological model explaining the chiral
symmetry breaking and the formation of the left-/right-handed
superhelical confostacks.

\subsubsection{Simple model for confoplex formation}

\begin{figure}[ptb]
\centering
\includegraphics[height=0.3\textheight]{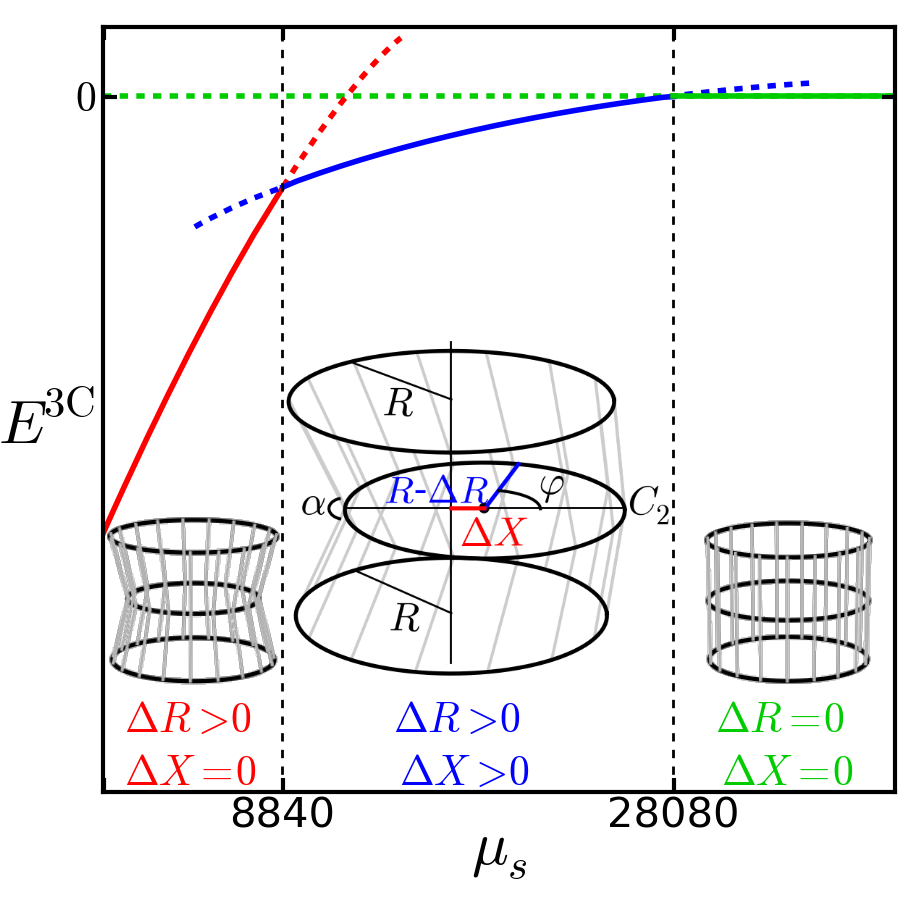}
\caption{
Elastic energy of a lattice section as a function of the stretching rigidity $\mu_s$ in the simple model.}%
\label{fig:energysimple}%
\end{figure}

In this section we revisit the transitions between the different
states of a single section (cylinder, partial and full confoplex)
and will explain them qualitatively. For all three states in
Fig.~\ref{fig:dimerring}, we observe that the horizontal bonds
form polygons which are close to circles lying in three parallel
planes. For a full catenoid-like confoplex, we see that the circle
in the middle (called $C_{2}$ because it is associated to the
anharmonic bonds $B_{2}$) has the smallest radius, whereas for the
partial confoplex there is additionally a small shift of the
center of $C_{2}$. This shift is in the direction of the
non-switched region of the wall, which thus bulges slightly
outwards. In view of these observations, it is tempting
to approximate a section of the tube by three circles that will
interact elastically in a very simplified manner (thus the results
can only be qualitatively compared with the simulation). By fixing
the center position and the radius $R$ of the lower and upper
circles, the only variables are now the position and the radius of
the circle $C_{2}$ in the middle. This circle can shift its center
horizontally by an amount of $\Delta X$. It can also decrease its
radius $R$ by $\Delta R$ with an energy cost
\begin{equation}
E_{s}=\frac{\mu_s}{26}(2\pi\Delta R)^{2}\; ,
\end{equation}
where $\mu_s$ is the
stretching rigidity of $C_{2}$. For a confoplex there is an
additional cost of bending energy due to the angle $\alpha$
between the lines joining the three circles at an azimuthal
position $\varphi$ of $C_{2}$ (see Fig.~\ref{fig:energysimple}).
This angle is chosen negative in the concave part of the confoplex and
given by $\sin\alpha\approx-2(\Delta R-\Delta X
\cos(\varphi))$ to first order in $\Delta R$ and $\Delta X$.
Similar to the anharmonic bending energy~(\ref{eq:anharmonic}) for the bonds $B_{2}$
used in the simulations, we write the bending energy of the simple
model as a double well potential and integrate it along
$C_{2}:$%
\begin{equation}
E_{b}=\oint_{C_{2}}\left(  A(\sin\alpha)^{4}+B(\sin\alpha)^{3}+C(\sin
\alpha)^{2}\right)  \text{d}\varphi\; .
\end{equation}
By minimizing the total energy of this three circle model,
$E^\text{3C}=E_{s}+E_{b}$, with respect to $\Delta X$ and $\Delta
R$, we observe the same three configurations as in the numerical
simulations (see Fig.~\ref{fig:energysimple} again): for
increasing $\mu_{s}$ the ground state shifts from a
full confoplex ($\Delta R>0$, $\Delta X=0)$ where $C_{2}$ can
easily stretch, to a partial one ($\Delta R>0$ and $\Delta X>0$) 
which is the result of a compromise between the stretching
and the anharmonic bending energy. For an even larger stretching
constant we finally obtain a cylinder ($\Delta R=\Delta X =0)$.
This simplified model which only comprises the stretching energy
and the anharmonic bending potential of the bond $B_{2}$ gives a
simple explanation for the very existence of confoplexes. Of
course, this reduced model is not capable to predict the precise
transition values; nonetheless it provides a qualitative
explanation for the observed morphologies in the simulation. Due
to its simplicity, it is highly intuitive and also analytically
tractable.


\subsubsection{Elastically-mediated
interactions between confoplexes and formation of confostacks\label{subsubsec:confostackformation}}

\begin{figure}[t]
\centering
\subfigure[][]{\label{fig:analogy1}{\includegraphics[height=0.11\textheight]{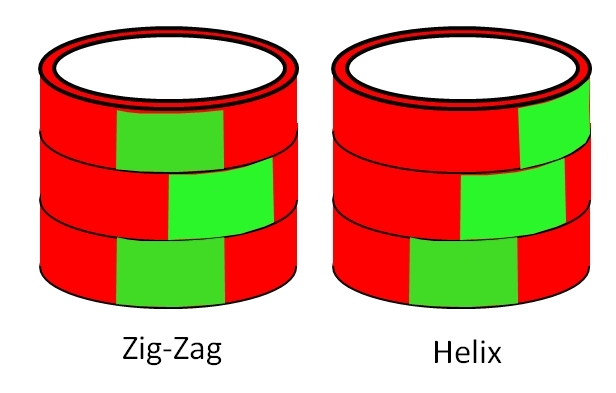}}}
\hspace{1cm}
\subfigure[][]{\label{fig:analogy2}{\includegraphics[height=0.11\textheight]{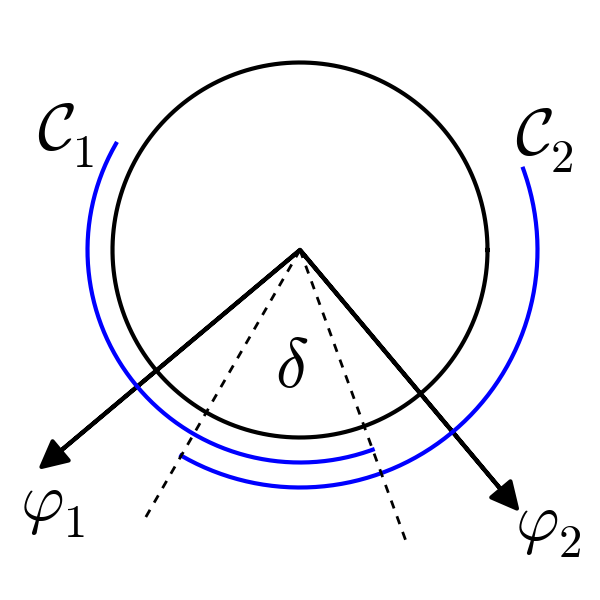}}}\newline%
\subfigure[][]{\label{fig:analogy3}{\includegraphics[width=0.45\textwidth]{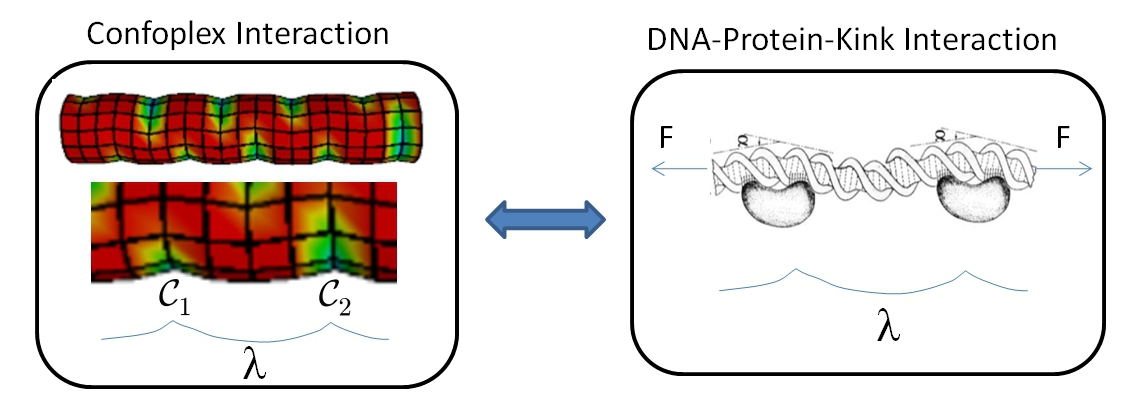}}}
\label{fig:analogy}\caption{(a) Possible quasiparticle
conformations for a short tube if the cooperative attractive
interaction and the elastic repulsion between neighboring partial
confoplexes is included into the model. (b) Variables used for
measuring the overlap of two confoplexes $\mathcal{C}_{1}$ and
$\mathcal{C}_{2}$. (c) Analogy between interacting confoplexes and
proteins bound to a semiflexible filament (picture adapted from
\cite{RudnickBruinsma99}, reprinted with permission from Elsevier), see text.
}%
\end{figure}

>From what we learnt previously from the simulations, the shape of
the confostack results from the competition between the
cooperative attractive interaction and the elastic repulsion
between partial confoplexes (see Fig.~\ref{fig:coop}). If both
interactions were short-ranged, one would observe the two
configurations depicted in Fig.~\ref{fig:analogy1} with equal
probability as they would have equal energies.

But the observed spontaneous symmetry breaking of the straight
tube, which leads to right- and left-handed superhelices, can only
be explained by the presence of an effective \textit{long-range}
repulsion between the confoplexes. Intuitively, the confoplexes
can be seen as quasiparticles which
deform the lattice surrounding them, generating an extended
elastic field. As a consequence the confoplexes will interact with
each other through long-range elastically-mediated repulsive
forces, extending further away than to the nearest neighbors.

A full analytical treatment of this interaction in our concrete
case is complicated by the presence of various lattice parameters
in the simulation, but a qualitative understanding is possible if
we retain only the dominant contributions. In this spirit,
consider two confoplexes $\mathcal{C}_{1}$ and $\mathcal{C}_{2}$
which azimuthally overlap with an angle $\delta$ (cf.
Fig.~\ref{fig:analogy2}) and interact only elastically without
cooperativity (left image of Fig.~\ref{fig:analogy3}). For
simplicity, let us assume that only the overlapping parts of
$\mathcal{C}_{1}$ and $\mathcal{C}_{2}$ interact strongly with
each other. This seems reasonable, since the elastic interaction
between the parts of the confoplexes which do not sit on the same
protofilament will be more screened around the tube. If the
confoplexes are shifted by an angle $\Delta\varphi=\left\vert
\varphi_{1}-\varphi _{2}\right\vert $ their interaction is then
proportional to the overlap angle $\delta$, and we can write
\begin{equation}
E_{\mathcal{C}_{1}-\mathcal{C}_{2}}\approx\frac{13\delta}{2\pi}E_{\mathcal{C}-\mathcal{C}}^{\text{single PF}}\; ,
\end{equation}
where $E_{\mathcal{C}-\mathcal{C}}^{\text{single PF}}$ is the
interaction energy of the parts of the two partial confoplexes
sitting on the same protofilament. For a small angular deviation
$\theta (u)$ with respect to the straight protofilament state this
energy can be approximated as
\begin{equation}
E_{\mathcal{C}-\mathcal{C}}^{\text{single PF}}=\frac{1}{2}\int\nolimits_{0}^{L}\left(
B_{\text{eff}}\left(  \frac{\text{d}\theta}{\text{d}u}\right)  ^{2}+F_{\text{eff}}\theta
^{2}\right)  du \; ,
\label{eq:energyPF}
\end{equation}
where $u$ is the arc length along the protofilament,
$B_{\text{eff}} \approx \mu_{b1}d$ is the bending rigidity of the
protofilament and $F_{\text{eff}} \approx {\mu}_{s}d$ is an intrinsic
effective tension due to the stretching rigidity of the bonds of
the protofilament.

Eq.~(\ref{eq:energyPF}) is formally equivalent to the interaction
energy of two proteins bound to the same side of a semiflexible
filament under an external pulling force. As shown in
\cite{RudnickBruinsma99} the elastic interaction is repulsive.
Similarly, cylindrical proteins bound to a locally flat membrane
are described with the same energy functional and repel as well
\cite{Weikl,Martin1,Martin2,MkrtchyanIngChen}). The range of this
repulsion is given by an elastic screening length
$\lambda=\sqrt{B_{\text{eff}}/F_{\text{eff}}}$ (see
Fig.~\ref{fig:analogy3}).

Despite the crude approximations considered here the basic physics of the interaction
is well captured.
In the presence of cooperativity, this long-range interaction leads to an interplay between nearest and
next-nearest neighbor repulsion between the quasiparticles; at short scale the
second configuration in Fig.~\ref{fig:analogy1} will be adopted. At a larger
scale the quasipolymer will thus form a helix on the lattice inducing a
superhelical structure of the tube in three-dimensional space.

Before embarking on the study of the dynamic properties of
polymorphic tubes, it is interesting to remark, that the
notion of quasiparticles, interacting by elastic fields in our
present case is rather similar to the notion of ``twist-kinks'', the
quasiparticles formed in squeezed helices confined to two
dimensions \cite{Squeezed helix}. In this context conformational
``twist-kink'' quasiparticles appear as a natural concept as well,
indicating a broader relevance of this perspective for prestressed
filaments with (hidden) internal degrees of freedom.


\section{Dynamics of clamped tubes: Confostacks diffuse through the lattice\label{sec:dynamics}}

The observation of microtubules clamped at one end has been a key
experiment revealing their anomalous behavior
\cite{pampaloni,Taute} and was the starting point for the
polymorphic tube model proposed in \cite{mt2010,mt2012}. In this
section we explore the dynamics of collective internal modes of
the lattice by simulating a tube of various lengths clamped at one
end, at finite temperature $T$. We focus on the simplest case in
which the tubes form circular arcs (helices with infinite pitch),
\textit{i.e.}, when they bear ideal untwisted confostacks with a
coherence length much larger than the tube's length.


\subsection{Numerical simulations}

\begin{figure}[t]
\centering
\subfigure[][]{\label{fig:rotaty}{\includegraphics[width=0.24\textwidth]{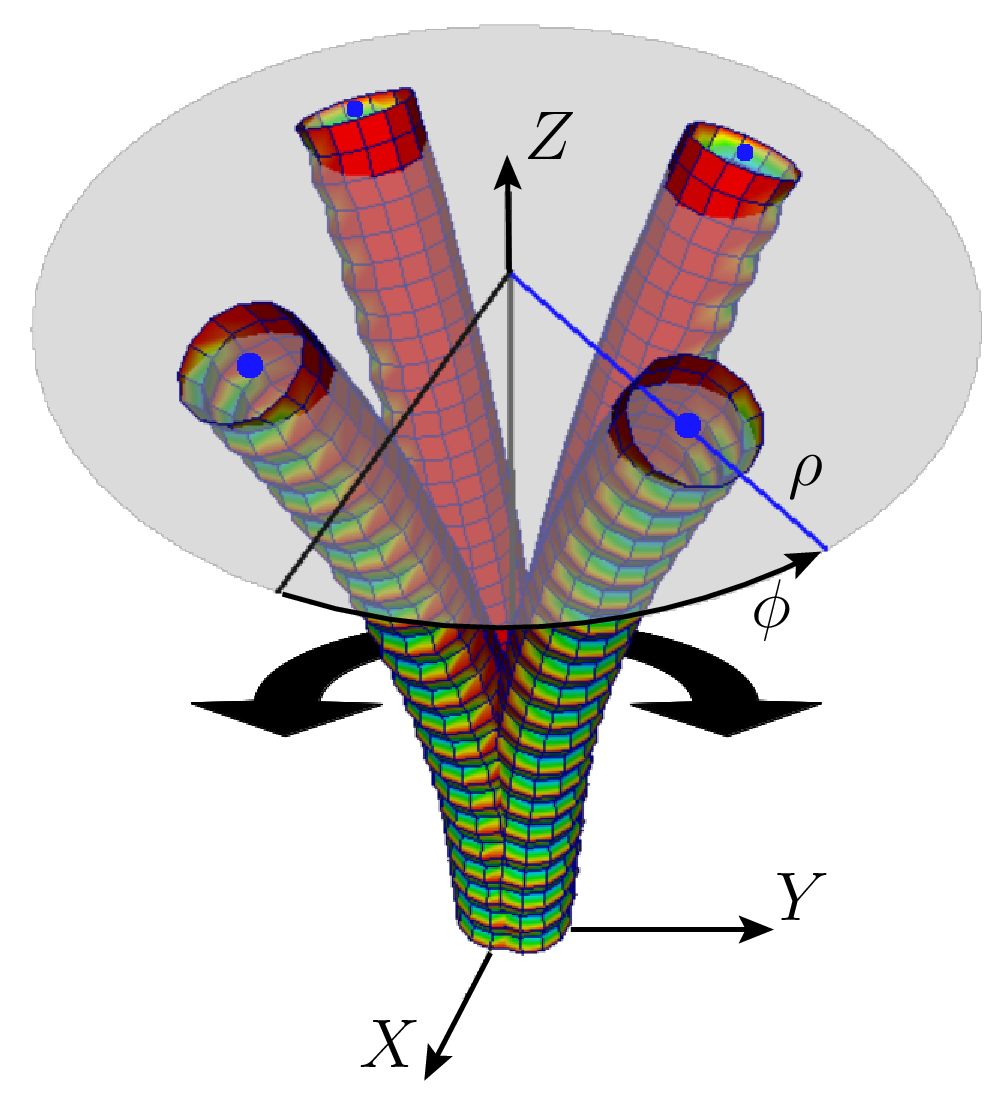}}}
\hspace{0.5cm}
\subfigure[][]{\label{fig:endpointdiffusion}{\raisebox{2.5cm}{\begin{minipage}{0.1\textwidth}\includegraphics[width=\textwidth]{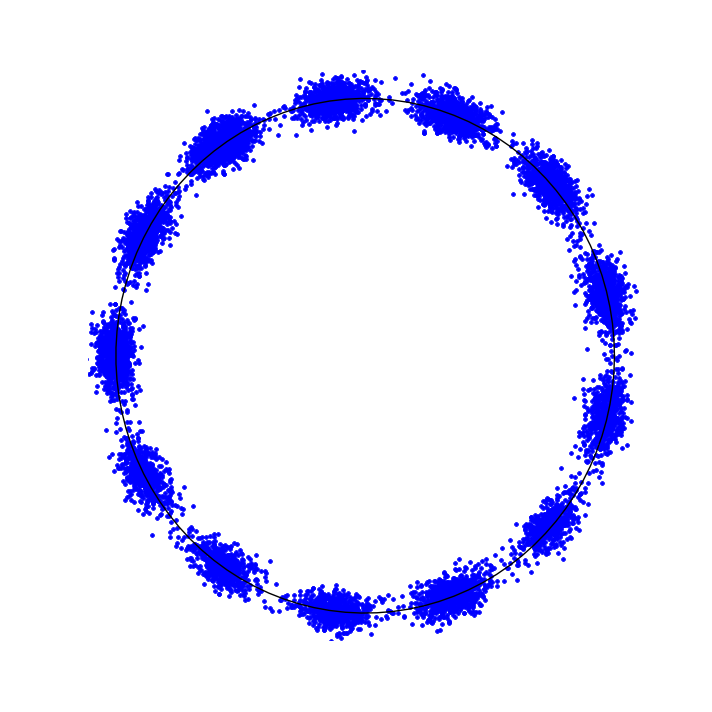}\\[-0.1cm] T=3
\\[0.3cm] \includegraphics[width=\textwidth]{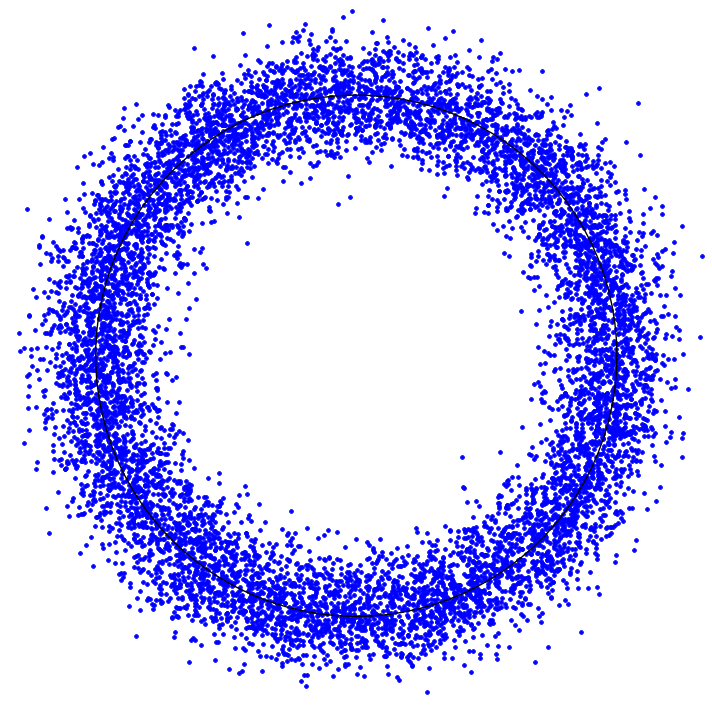}\\ T=10
\end{minipage}}}} \hspace{0.5cm}
\subfigure[][]{\label{fig:diffusion}{\includegraphics[width=0.5\textwidth]{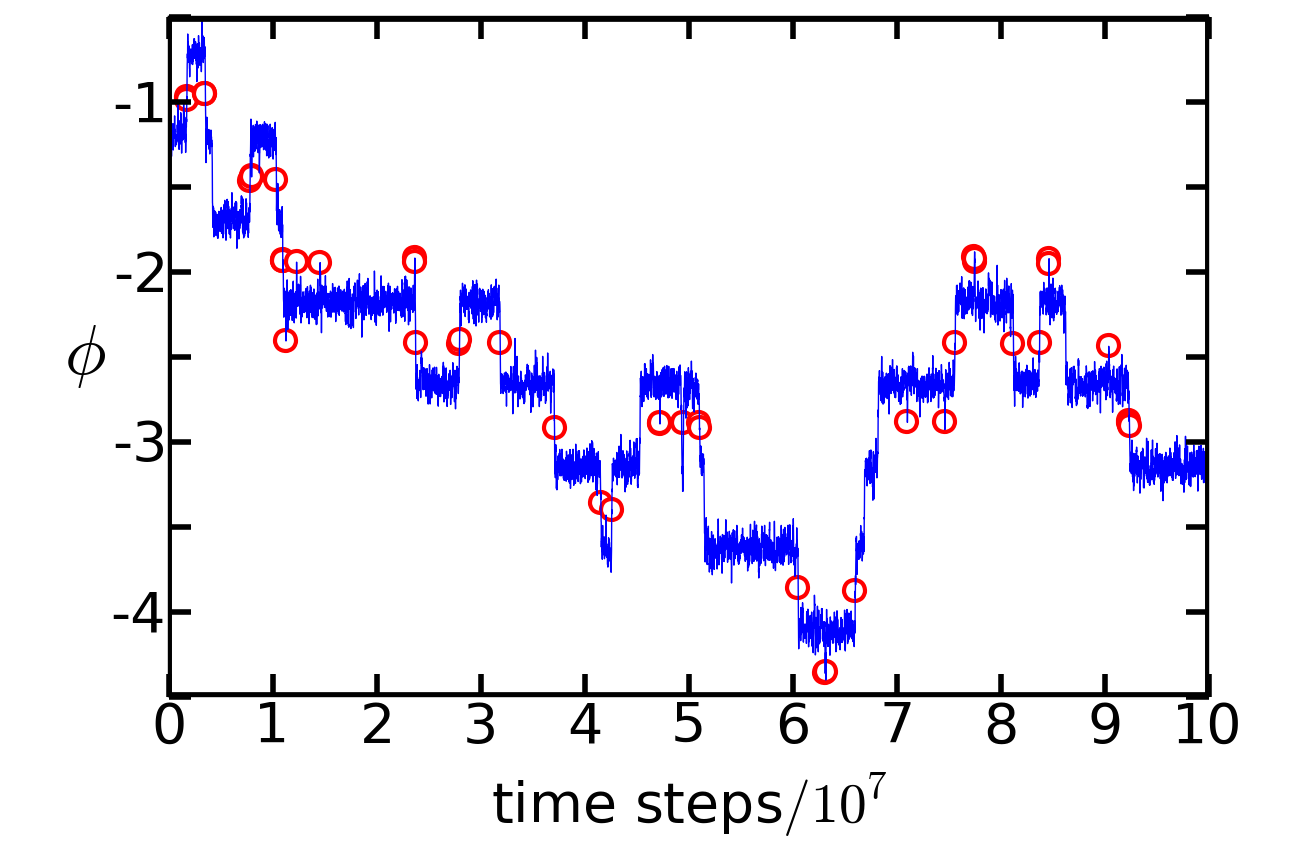}}}\caption{
(a) Rotating confostack clamped at the bottom end at finite
temperature $T$. The arc-shaped tube diffuses randomly around a
fixed axis. (b) Diffusion of the end point of the tube's
centerline for $T=3$ and $T=10$. For low temperatures the periodic
energy barrier $V(\phi)$ becomes apparent in the diffusion
pattern. (c) Temporal evolution of the azimuthal angle $\phi$ of
the end point of the tube's centerline at $T=3$. The red
circles highlight the transition regions in which the confostack
is at the maximum of the barrier (within $\pm0.05$ radian
from the location of this maximum).
}%
\label{fig:dynamics}%
\end{figure}

\subsubsection{Observation of the wobbling motion}

Even though the rigid attachment of the arc-shaped tube
seems to preclude large scale motion, the tube is not static in
shape. Instead, we clearly observe a random rotation of the tube around a fixed axis (see
Fig.~\ref{fig:rotaty}) in agreement with the predictions of the
wobbling motion for polymorphic tubes \cite{mt2010,mt2012}. It
is intuitively clear that the symmetry-broken
circular arc state can in principle explore all its equivalent
sister states, with the arc pointing in one of 13 possible
directions. However, it seems not obvious a priori if the
switching between the states can practically occur, and if so, on
which timescale. To quantify the dynamic behavior of this collective
``wobbling mode'', we measure the time
evolution of the centerline at the free end, which diffuses inside
an annular strip in the $\rho\phi$ plane (see Fig.~\ref{fig:endpointdiffusion}).

If the clamped tube was not polymorphic, but had the
shape of a \emph{static} circular arc instead, its elastic
fluctuations could be simply decomposed into a radial and an
azimuthal elastic mode. In this case we would expect a classical
``wormlike chain'' dynamic behavior, where the chain has merely an
intrinsic curvature. However, such an intrinsically curved elastic
filament would of course not rotate. In sharp contrast to the
wormlike chain case, the observed wobbling scenario allows an
efficient rotation (despite the fixed clamping of the end section)
 in our system. This rotary motion, which is the very signature of a
polymorphic lattice with broken symmetry, is apparently caused by a
confostack that moves azimuthally around the lattice.

Note that we have a peculiarly interesting ``polymer on a
polymer'' motif here: A quasipolymeric entity (the confostack)
exists and moves on the surface of a tube, that itself
can be considered as a polymer on larger scales. The
experimentally observable motion of the tube is slaved to the
motion of the confostack. The latter might be practically hidden
from direct experimental detection but it affects the tube's
centerline dynamics so severely that it becomes detectable by
tracing the tube's centerline.


\subsubsection{Energy barrier}
A closer look at the distribution of the azimuthal angle
$\phi$ of the tube's end reveals that the distribution is centered
around discrete angular values $\frac{2\pi n}{13},$
$n\in\mathbb{Z}$, corresponding to the $13$ protofilaments (c.f.
the blobs in Fig.~\ref{fig:endpointdiffusion}). This result
unveils the existence of a periodic potential $V(\phi)$---due to
the discrete lattice structure---in which both the
confostack and in turn also the free end of the tube diffuses.
Although the azimuthal rotation of the confostack can in principle
be continuous, it tends to visit the minimum of $V(\phi)$ much
more frequently. This potential is associated with an effective
energy barrier $\Delta E$ over which the confostack has to hop in
order to move azimuthally on the lattice from one minimum of
$V(\phi)$ to the next one. The consequences of $\Delta E$  on the
confostack movement will be revisited and explored later on.
\begin{figure}[t]
\centering
\includegraphics[width=0.45\textwidth]{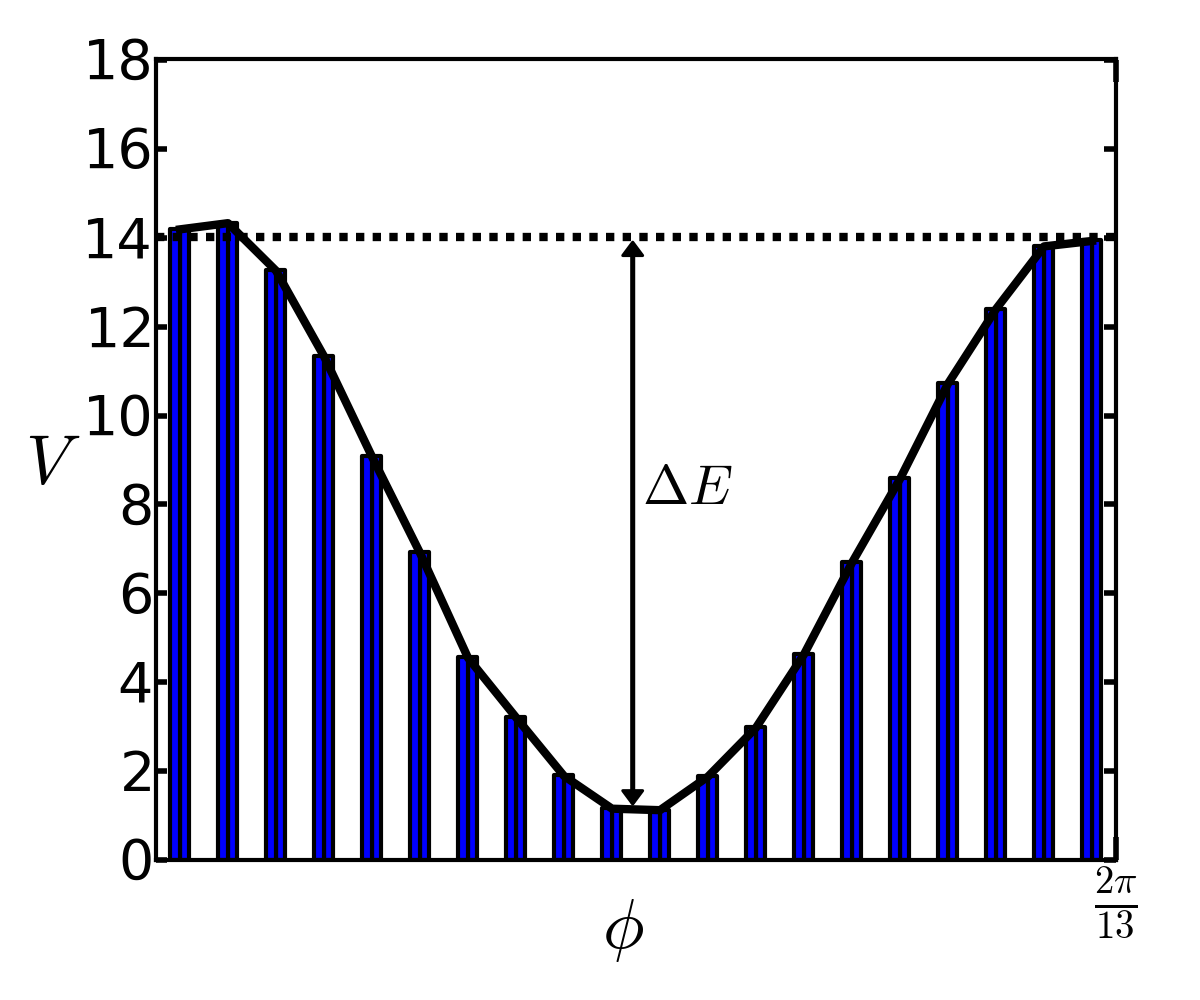}\caption{Histogram of the structural barrier $V(\phi)$ obtained
from the angular distribution function at $T=3$ and $N=20$.
}%
\label{fig:barrier}%
\end{figure}
Fig.~\ref{fig:barrier} shows a histogram of the logarithm of the
angular distribution function of the end of the centerline which
equals $-V(\phi)/(k_B T)$ modulo $\frac{2\pi}{13}$. This yields a
first estimate of the barrier $\Delta E$ measured from the
difference between the maximum and the minimum of $V(\phi)$:
$\Delta E \approx 12.5$. As we will see in
Secs.~\ref{subsubsec:tubedeflection} and
\ref{subsubsec:diffusionrevisited}, the same value will be found
with two other methods and is independent of the length $L$ of the
tube for all the lengths considered in the simulations.


\subsubsection{Rotary diffusion of the clamped tube\label{subsubsec:anomalousdiffusion}}

\begin{figure}[t]
\centering
\includegraphics[width=0.45\textwidth]{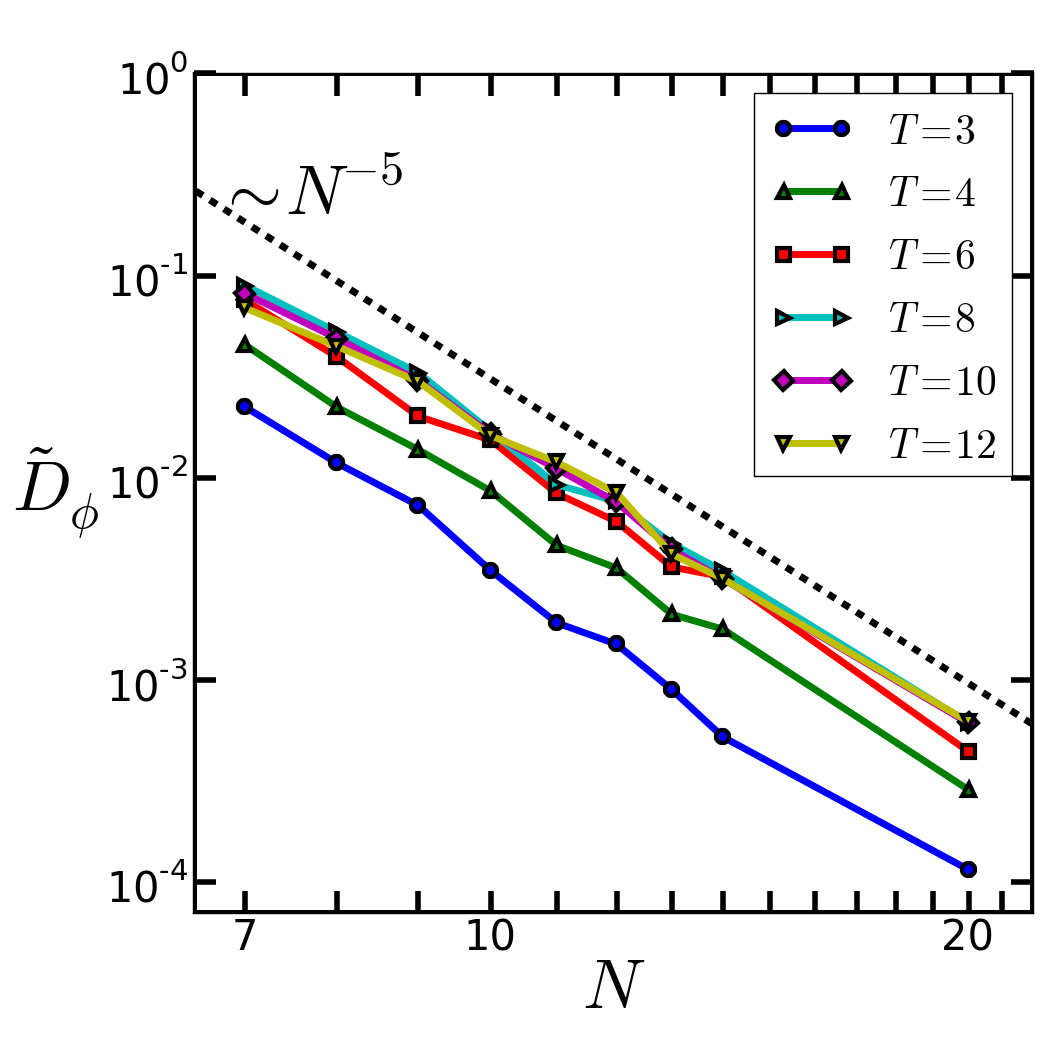}
\caption{Measured scaled diffusion coefficient (the mobility) of
the wobbling mode $\widetilde{D}_{\phi}=D_{\phi}/(k_{B}T)$ as a
function of the number of sections $N$ of the tube for different
temperatures T.
}%
\label{fig:dcoefunscaled}%
\end{figure}
>From the simulations one can get a more quantitative
understanding of the wobbling mode kinetics by measuring the mean
square displacement of the azimuthal angle $\phi$ of the end point
of the centerline of the tube. These measurements have been
performed for tubes of various lengths $L = 2Nd$ (with $N$ the
number of sections of the tube) and at different temperatures $T$.
The results display the typical behavior of a diffusive system
with a diffusion coefficient $D_{\phi}$: $\left\langle
(\phi(t+t_{0})-\phi(t_{0}))^{2}\right\rangle =2D_{\phi}t$. In
Fig.~\ref{fig:dcoefunscaled} we observe that the diffusion
coefficients normalized by the corresponding temperatures (the
mobility) $\widetilde{D}_{\phi}=D_{\phi}/(k_{B}T)$ all scale with
the length as $N^{-5}$ but do not collapse to a single
temperature-independent curve.

The scaling of the diffusion constant with length
$\widetilde{D}_{\phi}\sim N^{-5}$ is typical for a rotating rigid
circular arc moving through a fluid. The normalized diffusion
coefficient in this case is given by (see
Appendix~\ref{app:diffusion}) $\widetilde{D}_{\phi_{0}}=\frac{5}{8}%
\frac{N^{-5}}{\xi_{\perp}\kappa^{2}d^{5}}$, where $\kappa$ is the
curvature of the arc and $\xi_{\perp}$ the friction constant per
unit length.

The temperature dependence of $\widetilde{D}_{\phi}$ is
much more intriguing. Na\"ively, we would expect that, like for a 
rigid rotor moving through a fluid, $\widetilde{D}_{\phi}$ is
simply a constant with respect to the temperature. However, this
expectation is too simple as it considers only the friction of the
rotating arc through the external fluid medium and misses the
internal dynamics of the confostack, in particular the presence of
structural barriers, as discussed in the previous section. This
hidden internal dynamics nevertheless reflects itself in the
movement of the end point of the tube and leads to an additional
internal dissipation.

In summary, the simulations reveal that the tube's
friction must result from a combination of the external friction
of the arc in the fluid medium and a yet to be characterized
internal dissipation mechanism. In the next section, we explore
the origin of this inner dissipation mechanism. In particular, we
take a closer look at the conformational mode responsible for
these inner losses via barrier crossing.


\subsubsection{Barrier crossing transition state: The confostack-kink}

Until now, we have analyzed the wobbling motion by
tracking the time evolution of the end point of the centerline. In this
section we go a step further and take advantage of the simulation
to directly observe the shape of the confostack on the surface of
the lattice. At zero temperature, the tube assumes one of its
circular ground states oriented along one of the 13 possible
orientations. At any finite temperature, thermal excitations allow
for a continuous shape reorientation of the confostack between the
ground states by crossing the energy barrier $\Delta E$, while at
the same time reorienting the direction of the tube's curvature.

What is the critical confostack mode, \textit{i.e.}, the
conformation of the confostack at the top of the energy barrier
$\Delta E$?
%
\begin{figure}[t]
\centering
\subfigure[][]{\label{fig:kink1}{\includegraphics[width=0.45\textwidth]{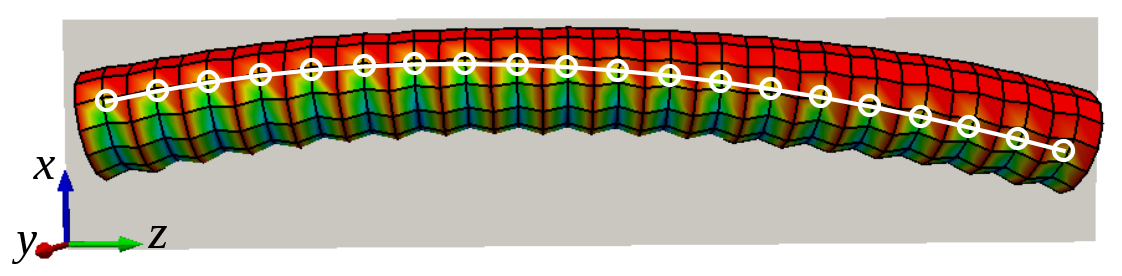}}}
\subfigure[][]{\label{fig:kink2}{\includegraphics[width=0.45\textwidth]{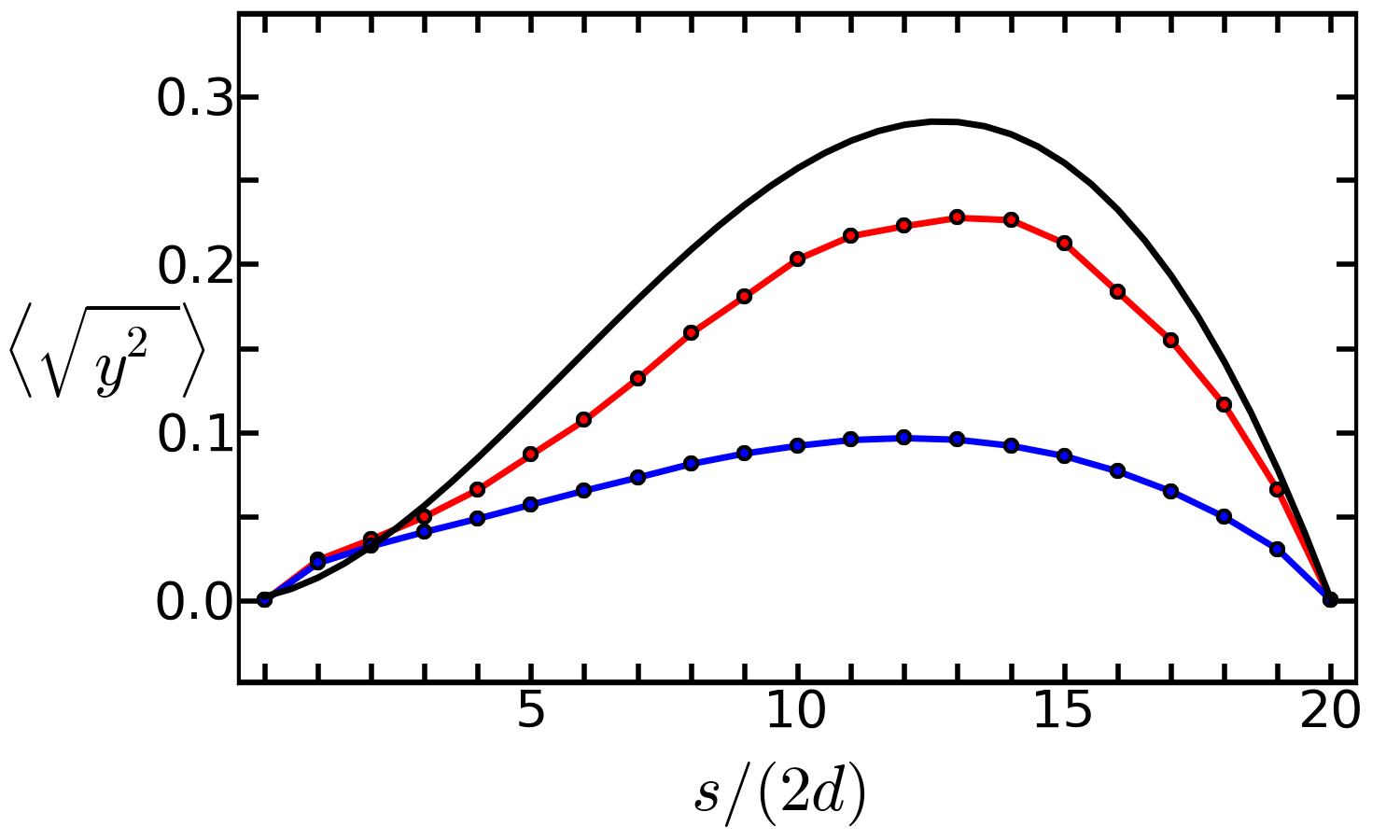}}}\caption{
(a) Simulation snapshot of a tube ($T=3$, $N=20$) in the
transition state regime. The tube is rotated into its natural
comoving frame in which the two ends of the tube's centerline lie
in the same plane $xz$. The shape of the confostack resembles a
kink which interpolates between two neighboring protofilaments.
This is the typical configurational mode of the transition for
large $N$. (b) Root mean square of the deflection $y(s)$ of all states 
(blue), the transition states (red), and the theoretical prediction for 
the transition states (black). }%
\end{figure}
%
A na\"{\i}ve first look into the noisy simulation snapshots of the tube in
space does not allow to identify the critical conformational mode on the
barrier (at the transition) in a simple way. But if one rotates the tube in
its natural co-moving frame of reference first (see next section for details),
one can observe the behavior of the confostack on the surface of the tube in
detail. This frame is defined by the $xz$ plane formed by the attachment point
and the end of the centerline of the tube.
Therefore, the deflection perpendicular to the $xz$ plane, $y(s)$,
satisfies the condition $y(0)=y(L)$, where $s$ is the arc length of the centerline.
At zero temperature, the confostack lies in the $xz$ plane and the
centerline of the tube is a circular arc. In this case the
co-moving and the laboratory frames are identical. At finite
temperature and for the transition regions, we typically observe a
confostack forming a kink on the lattice which will propagate
until the whole structure has crossed the barrier (see
Fig.~\ref{fig:kink1} for a snapshot). This confostack-kink
resembles a ``polymorphic'' dislocation that can be either left- or
right-handed, while reorienting the direction of the whole tube.

To capture and characterize the shape of the
confostack-kink under the conditions of strong thermal noise, we
have looked for statistical anomalies of the tube's shape in the
co-moving coordinates. Since the centerline of the tube leaves
the $xz$ plane only slightly, we expect that $x(s)$ still
describes a circular arc approximately
($x(s)\approx\frac{1}{2}\kappa s^{2}$). This is confirmed by the
numerical data (not shown). What turns out to be more interesting
is the $y$ direction. The root mean square $\left\langle
\sqrt{y^{2}(s)}\right\rangle _\text{Tr}$ of the deflection $y(s)$
of the centerline is computed over all transition states (red
circles in Fig.~\ref{fig:diffusion}). They correspond to the critical confostack-kink
at the top of the energy barrier $\Delta E$. This kink can either move to
the next minimum of the periodic potential $V(\phi)$ or return to the original
minimum. The curve $\left\langle
\sqrt{y^{2}(s)}\right\rangle _\text{Tr}$ is significantly
different from the measurement of the same quantity for all---transition
and non-transition---states of the simulation taken
together $\left\langle \sqrt{y^{2}(s)}\right\rangle_\text{All}$ as
shown in Fig.~\ref{fig:kink2}.

\subsection{Theory of the dynamics of the clamped tube}

To understand the unusual dynamics of the clamped tube
observed in the simulations and to interpret the behavior of the
confostack-kink at the barrier, in this section we make some
theoretical developments.

\subsubsection{Modelling a single confostack-kink}
We first define an external laboratory frame $(X,Y,Z)$. The tube
is clamped at the origin and is oriented in the $Z$ direction. For
small deviations around the $Z$ axis, the unit vector tangent to
the tube's centerline is approximately given by
$\boldsymbol{t}\approx(\theta_{X},\theta_{Y},1)$ with
$\theta_{X/Y}$ the deflection angles of the centerline in the
$X/Y$ direction. This deflection can be decomposed as the sum
$\theta_{X/Y}=\theta _{\text{el},X/Y}+\theta_{\text{pol},X/Y}$ of
a purely elastic deviation and a polymorphic one \cite{mt2012}.
The polymorphic contribution can be expressed in terms of a
polymorphic phase $\varphi\left(  s\right)  $ which stands for the
confostack's angular orientation with respect to the tube's
material frame. Therefore, one has
$\theta_{\text{pol},X}(s)=\kappa\int\nolimits_{0}^{s}\cos{\varphi(s^{\prime
})}\text{d}s^{\prime}$ and
$\theta_{\text{pol},Y}(s)=\kappa\int\nolimits_{0}^{s}\sin
{\varphi(s^{\prime})}\text{d}s^{\prime}$. Neglecting the purely
elastic fluctuations, the lateral displacement of the tube in the
$(X,Y)$ plane can be written as
\begin{eqnarray}
  X(s) & = & \int\nolimits_{0}^{s}\sin\left[  \theta_{\text{pol},X}(s') \right]  ds^{\prime} \; , \quad \text{and} \nonumber \\
  Y(s) & = & \int\nolimits_{0}^{s}\sin\left[ \theta_{\text{pol},Y}(s') \right]  \text{d}s^{\prime}
\; .
\end{eqnarray}
As defined previously, the co-moving frame ($x,y,z=Z)$ is given by the rotation
\begin{align}
x(s) &  =X(s)\cos\Phi+Y(s)\sin\Phi \; ,\nonumber\\
y(s) &  =-X(s)\sin\Phi+Y(s)\cos\Phi \label{Comv}%
\end{align}
with $\Phi=\arctan(Y(L)/X(L))$. Once we know the polymorphic phase
$\varphi\left(  s\right) $ of the
confostack configuration in the transition regime we can determine
the deflections $(x(s),y(s))$ of the centerline in the co-moving frame with the help of Eqs.~(\ref{Comv})
 and compare them with the experimental data.
To model the behavior of the confostack phenomenologically, we
assume that it moves along a tube of length $L=2Nd$ in a periodic
potential of amplitude $W$ with an effective polymorphic stiffness
$C_{p}$:
\begin{equation}
\Delta E (L)=\int_{-L/2}^{L/2}\text{d}s\;\left\{  \frac{C_{p}}{2}\varphi^{\prime
2}+\frac{W}{2}[1+\cos{(13\varphi)}]\right\}  \;,
\label{eq:barrier}%
\end{equation}
where the interval of the arc length $s$ of the centerline has been shifted for
mathematical convenience $s\in\left[  -L/2,L/2\right]  $.
The transition state is found by minimizing $\Delta E(L)$ with the
natural boundary conditions $\varphi^{\prime}(\pm L/2)=0$, \textit{i.e.},
no torque at the confostack ends. As shown in Appendix~\ref{app:confostackkink} there are two
regimes:

a) For $L<\pi\ell$ with $\ell=\sqrt{2C_{p}/(13^{2}W)}$, the minimal energy
barrier-crossing configuration is $\varphi\left(  s\right)  =0$. This
corresponds to a uniform rotation of the confostack as a block over the
barrier. In this case the barrier energy grows linearly with the length,
$\Delta E(L) = WL$.

b) For $L>\pi\ell$, a nontrivial barrier crossing solution
minimizing $\Delta E (L)$ is:
\begin{equation}
\varphi(s)=\frac{2}{13}\arcsin\left(\frac{1}{\sqrt{m}}\operatorname{sn}\left[\frac{s}{\ell
},1/m\right]\right) \; ,
\label{eq:confoshape}%
\end{equation}
where $\operatorname{sn}$ is the Jacobi sine function of parameter
$m>1$ \cite{Abra}. Eqn.~(\ref{eq:confoshape}) is a periodic
function in $s$. The solution which is monotonous and interpolates
between two successive minima of the periodic potential lies on a
finite interval given by the condition $L(m)=2\ell K[1/m]$, where
$K[1/m]$ is the complete elliptic integral of the first kind. In
the limit of large tube lengths $L$, we have $m\approx1$ and the
transition state confostack Eq.~(\ref{eq:confoshape}) becomes a
kink:
\begin{equation}
\varphi(s)=(4\arctan{(\text{e}^{\frac{s-s_{0}}{\ell}})}-\pi
)/13\label{confokink}%
\end{equation}
where $s_{0}$ is the position of the center of the kink on the lattice.

To determine the typical size of the confostack at the transition,
we look at the energy of the solution~(\ref{eq:confoshape})
computed in Appendix~\ref{app:confostackkink} (see Eq.~(\ref{deltaEapp})). In the
regime where the length $L\geq\pi\ell,$ the barrier $\Delta E (L)$
grows sublinearly and saturates at $\Delta E\approx0.05C_{p}/\ell$
for large $L$. As already mentioned, the notable observation from
the computer simulation is that $\Delta E\approx12.5$ for all the
lengths considered in the simulations, $L\geq14d$. Therefore,
$\ell\ll L$ and Eq.~(\ref{confokink}) is a good approximation for
the typical confostack configuration at the transition. This
analysis of a single confostack-kink allows us to determine the
theoretical root mean square of $y(s)$ at the transition which can
then be compared to the simulations.


\subsubsection{Deflection of the tube at the transition\label{subsubsec:tubedeflection}}


Assume that there is only a single confostack-kink of very short size $\ell\approx 0$ placed on a
random segment ($n=1,2,..N$) of the tube. Averaging over all positions of the kink with the
same probability we have computed $\left\langle \sqrt{y^{2}(s)}\right\rangle _{\text{model}}$
numerically for $N=20$ as shown in Fig.~\ref{fig:kink2}. We find that the model is close to the simulation
data of the transition state. This agreement is reassuring and validates the idea of
the confostack-kink as the main culprit for the crossing of the angular barrier.

In addition, from the ratio of  $\left\langle
\sqrt{y^{2}(s)}\right\rangle $ in this model over the simulation
data for all frames (transition and non-transition states) we can
infer the kink density in the simulation. Indeed
the probability to find the kink anywhere is $P\approx$ $\frac{\text{max}%
\left(\left\langle \sqrt{y^{2}(s)}\right\rangle_\text{All}\right)}{\text{max}\left(\left\langle
\sqrt{y^{2}(s)}\right\rangle _{\text{model}}\right)}\approx0.368$.
Therefore, the kinks are roughly three times more frequent in the transition
states as expected from a kink-mediated barrier crossing mechanism.

This result allows us to determine $\Delta E$ in another way:
The probability of finding a single kink on any segment of the tube ($n=1,2,..N$)
is given by:
\begin{equation}
P_{1}=\frac{\exp(-\frac{\Delta E}{k_{B}T})}{1+\exp(-\frac{\Delta E}{k_{B}T})}
\; .
\end{equation}
The probability to find at least one kink on the tube is then
given by $P=1-(1-P_{1})^{N}$. For a tube with $N=20$ segments at
$T=3$ we find that $P\approx 0.368$ leads to an energy barrier
$\Delta E\approx 12.5$ in agreement with the simulations. The
corresponding $P_{1}\approx 0.018$ shows that for $N=20$ the
density of kinks is $NP_{1}\approx 0.37$. Since the kinks are
roughly three times more frequent at the transition we deduce that
the density of kinks at the transition is about one which
justifies our original assumption.


\subsubsection{Rotor diffusion revisited\label{subsubsec:diffusionrevisited}}

The theoretical modelling in the previous sections allowed us to deepen our understanding
of the confostack dynamics in the presence of a barrier. With this knowledge we can return
to the anomalous diffusion of the clamped tube.
Approximating the internal periodic barrier $V(\phi)$
by a sinusoidal of the form $V(\phi)=(\Delta E/2)\cos(13\phi)$, the azimuthal
diffusion coefficient $D_{\phi}$ of the free end of the centerline should obey
the relation \cite{Risken}
\begin{equation}
D_{\phi}=D_{\phi_{0}}\left[  I_{0}\left(  \frac{\Delta E}{2k_{B}T}\right)
\right]  ^{-2}\;,\label{eq:diffusion}%
\end{equation}
where $I_{0}$ denotes the modified Bessel function of the first kind \cite{Abra}. The
robustness of the scaling $\sim N^{-5}$ of $D_{\phi}$ confirms again that the energy
barrier $\Delta E$ has to be independent of $N$ for the lengths we considered.
To compare the diffusion at different temperatures, we introduce a
new scaled (temperature-independent) diffusion coefficient $\tilde{D}%
=\frac{D_{\phi}}{k_{B}T}\left[  I_{0}\left(  \frac{\Delta E}{2k_{B}T}\right)
\right]  ^{2}$.

\begin{figure}[t]
\centering
\includegraphics[width=0.45\textwidth]{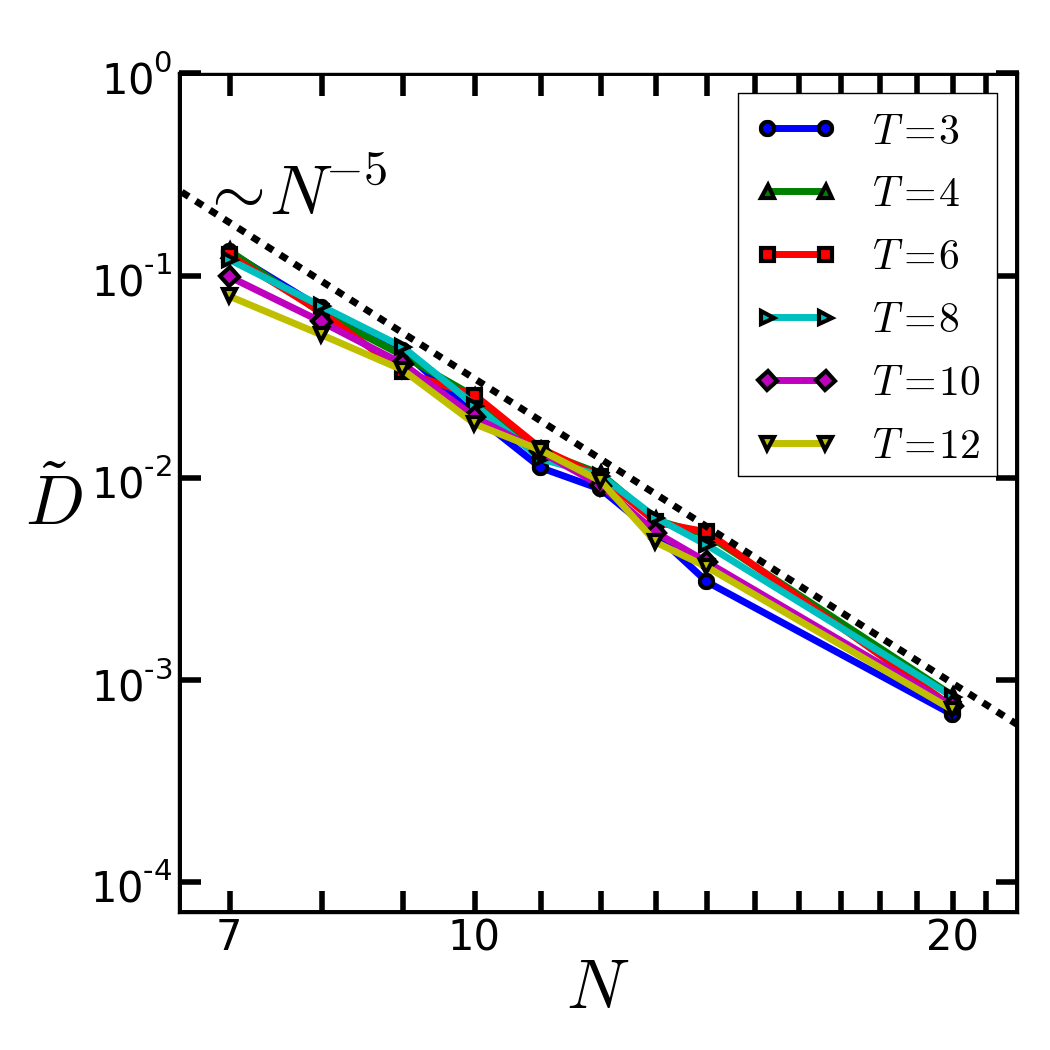}
\caption{Scaled diffusion coefficient $\tilde{D}$ as a function of the number
of sections $N$ of the tube for different temperatures. The $N^{-5}$ scaling
can be explained by the diffusion of a clamped circular arc in a fluid. The
collapse of the data on one master curve can only be understood when the
internal polymorphic dynamics is properly taken into account (see text).
}%
\label{fig:dcoef}%
\end{figure}
In Fig.~\ref{fig:dcoef} we see that the simulation results for
different temperatures collapse to a single curve for $\Delta
E\approx12.5$. This coincides again with the previous results. The
master curve is slightly below the curve for the perfect rotor
diffusion coefficient $\widetilde{D}_{\phi_{0}}$.

This small discrepancy can be understood as the
polymorphic tube is softer than the ideal rigid rotor. Indeed, it
has to sustain some additional deformations due to the presence
and migration of defects in the confostack. The reaction
coordinate between two angular orientations is therefore in
reality not straight as it would be for a rigid rotor, but instead
the system has to take a non-straight ``detour'' in the
configuration space. This leads to an apparent reduction of the
diffusion constant along the ideal (shortest path) rotary reaction
coordinate.

To summarize, we have seen that both the external friction of the
arc (through the external fluid medium) as well as an internal
dissipation mechanism (crossing an inner energetic barrier) of the
tube combine together into an effective friction $\xi_{\phi}=k_BT
/D_{\phi}$. Note that contributions of an anomalous friction in
the short modes of microtubules were reported by Jansen \&
Dogterom \cite{Jansen}, Taute et al. \cite{Taute} and Brangwynne
et al. \cite{Brangwynne}. It was speculated by these researchers
that some form of internal dissipation mechanism was at work. In
this section we have seen how internal barriers and conformational
cooperativity give rise to such internal friction phenomena.


\section{Conclusion}

The elastic and thermal properties of isotropic and anisotropic
macromolecular tubes have been the focus of scientific research
for decades. In this paper we have studied a new system consisting
of a tubular lattice whose individual elements can switch between
a flat and a curved configuration. This triggers the birth of
\textquotedblleft confoplexes\textquotedblright, conformational
quasiparticles that interact \textit{via} long-range repulsive
interactions mediated by the elasticity of the lattice. By
introducing structural cooperativity (as motivated by biological
systems) and in turn \textquotedblleft
polymerizing\textquotedblright\ a number of confoplexes on the
tubular lattice, a plethora of different phenomena have been
discovered: the tube spontaneously breaks its cylindrical symmetry
and forms superhelical structures in three-dimensional space.
Remarkably, at finite temperature, the movement of the
quasipolymer built out of confoplexes on the lattice constantly
reshapes the whole tube inducing a random rotation for a clamped
tube. This dynamics has been studied in detail by numerical
simulations and phenomenological theory. We found that the
quasipolymer---the confostack---has to cross a periodic energy
barrier to move azimuthally on the lattice. We observe that the
typical conformational mode for barrier crossing is a conformational
defect that we termed a confostack-kink. The associated kink-dynamics
of the confostack on the lattice was the clue to explain
the behavior of the diffusion coefficient of the clamped tube.

Looking back at what we have learned from the present tube model, we note one
interesting perspective crystallizing out. It is the idea of localized
conformational quasiparticles living on the lattice. We have seen how an
elementary quasiparticle---the confoplex---emerges and how it interacts with others of
the same kind via elastic lattice modes. Once an additional
cooperative interaction is introduced, the particle-like confoplexes are
forced together into an extended polymer-like conformational object---the
confostack. However, curiously the quasipolymeric confostack tends, once
again, to decompose into smaller discrete entities. This gives rise to another
discrete localized particle-like entity---the confostack-kink.


Exploring the implications of the quasiparticle point of view to study the
``confotronics'' of a multitude of concrete biological (tubular or cylindrical)
monomer lattices, like flagellin, microtubules and actin, promises quite some
excitement ahead.



\begin{acknowledgments}
The PMMS (P\^{o}le Messin de Mod\'{e}lisation et de Simulation) is
acknowledged for providing the computer time. We thank Albert Johner,
Jean-Fran\c{c}ois Joanny, Carlos Marques, Helmut Schiessel, Ren\'{e} Messina
and Norbert Stoop for stimulating discussions.
\end{acknowledgments}


\appendix

\section{Details of the simulations\label{app:simulations}}

Each vertex of the lattice is treated as a bead subject to the
equation of motion:
\begin{align}
m \ddot{\mathbf{r}} = \mathbf{f} - m\gamma\dot{\mathbf{r}} + \mathbf{\Gamma}%
\; ,
\end{align}
where $m$ is the mass of the bead, and $\gamma$ is the damping
constant. $\mathbf{\Gamma}$ denotes the Gaussian white noise, and
$\mathbf{f}$ is the sum of the (elastic) forces acting on the
bead.

The corresponding discrete velocity Verlet algorithm reads
\cite{Schlick}
\begin{align}
\dot{\mathbf{r}}_{n+1/2}  &  = \dot{\mathbf{r}}_{n} +
(\mathbf{f}(\vec{r}_{n}) - m \gamma\dot{\mathbf{r}}_{n} +
\mathbf{\Gamma}_{n}) \frac{\Delta t}{2
m}
\; ,
\nonumber\\
\mathbf{r}_{n+1}  &  = \mathbf{r}_{n} + \dot{\mathbf{r}}_{n+1/2} \Delta t
\; ,\\
\dot{\mathbf{r}}_{n+1}  &  = \dot{\mathbf{r}}_{n+1/2} +
(\mathbf{f}(\vec {r}_{n+1}) - m \gamma\dot{\mathbf{r}}_{n+1} +
\mathbf{\Gamma}_{n+1}) \frac{\Delta t}{2 m}\nonumber
\; ,
\end{align}
where the noise term is sampled with zero mean and a variance
$\langle \Gamma_{i}^{2} \rangle= 2 k_{B} T_0 m \gamma/ \Delta t$
for each component $i$ with room temperature $T_0$.
In the simulations, we set the Boltzmann constant, the mass and the
damping coefficient to unity. The integration time step is set to
$\delta t = 0.001 \, \tau$, where $\tau$ denotes the unit of time
in the simulations.

The elastic parameters in the simulations are chosen in the following manner:
The rest lengths are set to $d^{(0)}_1=d^{(0)}_2=d^{(0)}_3=d$ and
$d^{(0)}_4=\sqrt{2}\,d$, where $d=4\,$nm is the size of a monomer, chosen
as the unit length of the system.
The angular constants are given by $s^{(0)}_1=s^{(0)}_4$ and $s^{(0)}_3=\sin{\frac{2\pi}{13}}$.
These values ensure that the lattice is cylindrical in the absence of the anharmonic potential.

The value of the stretching rigidity $\mu_{s}$ is specified in the
main text. For the diagonal bonds, we apply $\mu_{s4} = 1000 \,
k_{B} T_0 / d^{2}$ which is sufficient to conserve the rectangular
nature of the subunits. A bending rigidity $\mu_{b4} = 500 \,\,
k_{B} T_0 $ is chosen for the diagonal bonds. The bending rigidity
of the bonds $B_{3}$ is set to $\mu_{b3}=1500 \,\, k_{B} T_0 $.

The coefficients of the anharmonic bending potential
$E_2^{\text{BAH}}$ of the bonds of type $B_{2}$, given in Eq.~(\ref{eq:anharmonic}),
are set to $A= 3350\,k_{B} T_0$, $B = 2960\,k_{B} T_0$, $C =
290\,k_{B} T_0$ to ensure that $\Delta G = 200 \,\,k_{B} T_0$ and
$\delta G = 1 \,\, k_{B} T_0$. The global minimum of
$E_2^{\text{BAH}}$ is situated at $\bar{s}_{2} = -0.588$. This
value leads to a preferred curved state with an angle $36^{\circ}$
of the free dimer. Note that, if we approximated the double well
potential by a harmonic one around $\bar{s}_{2}$, we could write
$E_2^{\text{BAH}} \approx \frac{\mu_{b2}}{2}
\sum_{\{s_2\}}\left(s_2-\bar{s}_{2} \right)^2$ with an effective
bending rigidity $\mu_{b2}\approx 2000 \,k_{B} T_0 $, a value
slightly larger than $\mu_{b3}$.


\section{Rigid rotor dynamics in a surrounding fluid \label{app:diffusion}}

To explain the temporal diffusion of the clamped polymorphic tube,
we need the diffusion coefficient of a rotating rigid circular arc
in a fluid. Assuming a constant friction per unit length for the
cross section of the arc, $\xi_{\perp}$, the Rayleigh dissipation
functional is given as
$P_{\text{diss}}=\frac{1}{2}\xi_{\perp}\int_0^L\dot{\rho}(s,t)^{2}\text{ds}=\frac{1}%
{2}\xi_{\perp}\int_0^L\rho^{2}\omega^{2}\text{ds}$ where $\omega$
is the angular velocity of the rotor. Since the deflection of the
arc is $\rho(s)=\frac{1}{2}\kappa s^{2}$, we obtain
$P_{\text{diss}}=\frac{1}{2}\xi_{\phi_{0}}\omega^{2}$ with an
effective rotational friction $\xi_{\phi_{0}}$ that can be
directly read off
$\xi_{\phi_{0}}=\frac{\xi_{\perp}\kappa^{2}L^{5}}{20}$. The
azimuthal diffusion coefficient
$D_{\phi_{0}}=\frac{k_{B}T}{\xi_{\phi_{0}}}$ of the free end of
the centerline is thus:
\begin{equation}
D_{\phi_{0}}=\frac{5}{8}\frac{k_{B}T}{\xi_{\perp}\kappa^{2}d^{5}}N^{-5}
\label{diffusion}%
\end{equation}
where $N=L/(2d)$ (see Sec.~\ref{subsubsec:anomalousdiffusion}).


\section{The emergence of a polymorphic confostack-kink\label{app:confostackkink}}

The wobbling mode of the clamped polymorphic tube is due to the
formation of a particular confostack configuration which allows
the tube to cross the angular energy barrier. The details of this
movement can be understood with the help of a phenomenological
model.

The energy of a confostack, Eq.~(\ref{eq:barrier}), can be written
in terms of the transformed angle $\psi=13\varphi$ and the scaled
polymorphic stiffness $\widetilde {C}_{p}=13^{-2}C_{p}$ as
\begin{equation}
\Delta E(L) = \int_{-L/2}^{L/2}\text{d}s\;\left[ \frac{\widetilde{C}_{p}}{2}%
\psi^{\prime2}  +\frac{W}{2}(1+\cos{\psi})\right]  \;.
\label{deltaEappendx}%
\end{equation}
This energy has to be minimized with the boundary conditions
$\psi^{\prime}(\pm L/2)=0$ (no torque) to find the
barrier-crossing configuration of the confostack. This gives the
Euler-Lagrange equation
\begin{equation}
  \ell^{2}\psi^{\prime\prime}=-\sin\psi
\end{equation}
with $\ell=\sqrt{2\widetilde{C}_{p}/W}$. There is always the
trivial configuration $\psi\left(  s\right)  =0$ with energy
$\Delta E (L)=WL$ and a non-trival antisymmetric solution with
$\psi\left(  0\right)  =0$:
\begin{equation}
\psi(s)=2
\operatorname{am}\left(\frac{s}{\ell\sqrt{m}},m\right)\text{ \
with }m=\frac{4}{2+C}
\label{psyapp}%
\end{equation}
with $\operatorname{am}(s,m)$ the Jacobi amplitude function of
parameter $m\in\left[ 0,1\right]$ \cite{Abra} and $C$ a constant
of integration. The solution describes a revolving pendulum wich
is a multi-kink solution. A single kink can be defined on a finite
region of size $L$, given implicitly by the
relation $\sqrt{m}K[m]=L/2\ell$. However, the boundary conditions
$\psi^{\prime}\left(  \pm L/2\right) =0$ can never be satisfied in
this case, except for $L\rightarrow\infty$ $(m\to 1)$.
Nevertheless, a physical solution for a finite
length of the confostack can be obtained by analytic continuation of Eq.~(\ref{psyapp}) choosing $m>1:$%
\begin{equation}
\psi(s)=2\arcsin\left(\frac{1}{\sqrt{m}}\operatorname{sn}\big[\frac{s}{\ell},1/m\big]\right)\text{
\ \ with }m>1 \; .
\label{Confoapp}%
\end{equation}
This solution is a periodic function. It is monotonous on a finite
interval given by $L(m)=2\ell K[1/m]$. The associated energy is
\begin{equation}
\Delta E (m)=\frac{\widetilde{C}_{p}}{\ell}\left(
8\operatorname{E}[1/m]-\frac{4(m-1)}{m}
\operatorname{K}[1/m]\right) \; .
\label{deltaEapp}%
\end{equation}
with $\operatorname{K}$ and $\operatorname{E}$ being the complete
elliptic integrals of the first and the second kind.

For $m\simeq1$, and thus $L$ very large,
$\psi(s)=4\arctan{(\text{e}^{\frac{s}{\ell}})}-\pi$ and the
barrier energy is a constant $\Delta
E\approx8\widetilde{C}_{p}/\ell$. Increasing $m$ decreases $L,$
and the energy stays close to its plateau value
$8\widetilde{C}_{p}/\ell$. When $L$ approaches
 $L_{c}=\pi\ell$ , $\Delta E$ decreases sublinearly and reaches $\Delta E\approx2\widetilde{C}%
_{p}\pi/\ell$ for $L=L_c$ $(m=\infty)$. For $L<L_{c}$, $\psi\left(
s\right) =0 $ is the only solution, and the barrier scales
linearly with $L$ in this regime. A comparison with the results of the numerical
simulations is presented in the main text.


\end{document}